\documentclass[aps,prd,showpacs,reprint]{revtex4-1}
\usepackage{amsmath,amsthm,latexsym,amssymb,amsfonts,epsfig}

\allowdisplaybreaks

\newcommand{\expec}[1]{\left\langle #1 \right\rangle}
\newcommand{\bra}[1]{\left\langle #1 \right\rvert}
\newcommand{\ket}[1]{\left\lvert #1 \right\rangle}

\newcommand{\at}{\text{at}}
\newcommand{\ma}{\text{max}}
\newcommand{\mi}{\text{min}}
\newcommand{\lp}{l_\text{P}}
\newcommand{\sch}{\text{S}}

\DeclareMathOperator{\tr}{tr}

\begin{document}
\date{15. Feb 2011}

\title{Energy equipartition and minimal radius in entropic gravity}

\author{Hanno Sahlmann}
\email{sahlmann@apctp.org}
\affiliation{Asia Pacific Center for Theoretical Physics, Pohang (Korea)}
\affiliation{Physics Department, Pohang University of Science and Technology, Pohang 
	(South Korea)}

\preprint{APCTP Pre2011-003}

\begin{abstract}
In this article, we investigate the assumption of equipartition of energy in arguments for the entropic nature of gravity. It has already been pointed out by other authors that equipartition is not valid for low temperatures. Here we additionally point out that it is similarly not valid for systems with bounded energy. Many explanations for black hole entropy suggest that the microscopic systems responsible have a finite dimensional state space, and thus finite maximum energy. Assuming this to be the case leads to drastic corrections to Newton's law for high gravitational fields, and, in particular, to a singularity in acceleration at finite radius away from a point mass. This is suggestive of the physics at the Schwarzschild radius. We show, however, that the location of the singularity scales differently. 
\end{abstract}

\maketitle
\section{Introduction}
In the interesting recent article \cite{Verlinde:2010hp}, Verlinde argued that the gravitational force is an entropic force, and thus emergent from some more fundamental theory. This is a continuation of the quest to explain gravitation as the thermodynamic
limit of some underlying microphysics, begun in earnest in the seminal work \cite{Jacobson:1995ab} (see also \cite{Padmanabhan:2002xm}). While there is an ever-growing literature on extensions and applications of these ideas, there are also serious concerns regarding the theoretical \cite{Visser:2011jp} and experimental \cite{Kobakhidze:2010mn,Chaichian:2011xc,Kobakhidze:2011gr,Chaichian:2011hc} viability of Verlinde's proposal. This debate is far from settled, and entropic explanations of the gravitational interaction remain controversial.

Verlinde assumes that gravity admits a holographic description, in terms of systems that exist on equipotential surfaces and represent ``bits of information.'' Each such system is assumed to contribute 
an area $l^2_\at$, and thus 
\begin{equation}
N_\at=\frac{A}{l^2_\at}
\label{eq_nat}
\end{equation}
is the integer number of systems that make up an equipotential surface of area $A$. 
We will call these systems ``atoms of area.'' In \cite{Verlinde:2010hp}, $l_\at$ is equal to the Planck length $\lp$, but the general argument can support an independent length scale \cite{Klinkhamer:2010qa}, or even several atom species with different sizes \cite{Sahlmann:2010jh}. 

\begin{figure}
\centerline{\epsfig{file=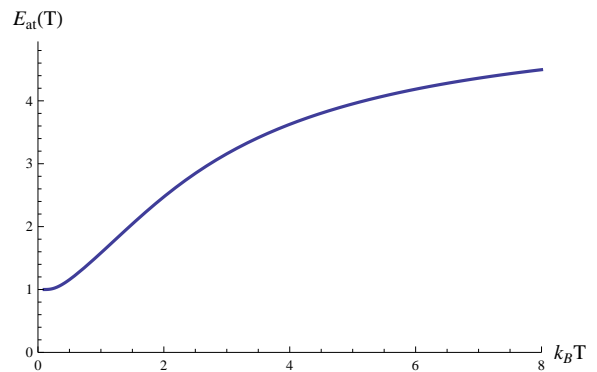,scale=0.6}}
\caption{\label{fi_generic} Energy versus temperature for the system described in Sec.\ \ref{se_mod} (in units of the energy spacing $E_0$, with $N=10$ energy levels).}
\end{figure}

A key point in the argument is the use of the energy equipartition law
\begin{equation}
E_\at=\frac{1}{2}k_BT.
\label{eq_equi1}
\end{equation}
Here, $E_\at$ is the average energy of a single atom of area, and $T$ the temperature corresponding to the gravitational acceleration $a$ of a test mass via the Unruh effect, 
\begin{equation}
T=\frac{1}{2\pi}\frac{\hbar}{ck_B} a. 
\label{eq_unruh}
\end{equation} 
We also refer the reader to the interesting independent discussion of this law in the context of gravitational physics in \cite{Padmanabhan:2010xh}. 
It has been pointed out in the literature \cite{Gao:2010fw,Kiselev:2010xi} that this formula is usually valid only in the high temperature regime, with corrections due to the quantization of energy expected at low temperatures. This is interesting because the temperatures associated to normal gravitational accelerations are extremely low. Here, we would like to make a complementary point: There are also corrections if we consider atoms with \emph{energy bounded from above}. In this case the average energy carried by 
an individual atom starts to saturate when $k_\text{B}T$ becomes comparable to the highest energy level of the atom, and corrections to \eqref{eq_equi1} ensue, see for example Fig.\ \ref{fi_generic}.

One reason for the assumption of finite energy comes from explanations for black hole entropy. In many of them, the systems that account for the entropy are finite dimensional, at least as long as the area of the black hole is held fixed. 
It is natural to identify black hole entropy with the entropy from the entropic gravity scenario for the special case that the equipotential surface is a black hole horizon. This suggests that the atoms of area have a finite dimensional state space in general, and hence bounded energy.  

If the atoms indeed have bounded energy, there is a maximum total energy that a system of $N_\at$ atoms can carry. We will see that in the gravitational context, this translates to a \emph{minimal radius} $R_\mi$ -- for a given mass $M$ -- above which the gravitational acceleration can be calculated. At the minimal radius, the gravitational acceleration diverges. This is strongly reminiscent of the physical situation at the Schwarzschild radius $R_\sch$. A straightforward identification of $R_\mi$ with $R_\sch$ is however prohibited by the scaling of the former: We find 
\begin{equation}
R_\mi = \sqrt{\frac{\alpha}{2\pi}} \sqrt{R_\sch R_\at}
\end{equation}
where $R_\at= c \hbar/E_\ma$ is the (reduced) Compton wavelength of an atom at maximum energy, and $\alpha$ is a numerical constant that depends on the detailed physics of the atoms. Moreover, there is an argument that shows that $R_\mi$ must always stay below $R_\sch$, see the note added at the end of Sec.\ \ref{se_conc}. Thus the minimal radius is only comparable to the Schwarzschild radius, if $R_\at$ and $R_\sch$ are comparable, in other words
$E_\ma\approx E_\text{P} (m_{\text{P}}/M)$. This is for example the case for $M$ of the order Planck mass, and Planck energy atoms. For larger masses, and the same $E_\ma$, $R_\mi$ stays far below the Schwarzschild radius, and Verlinde's argument reproduces Newton's law beyond $R_\sch$. We should stress that this result is obtained by working with the nonrelativistic version of Verlinde's argument. Since the corrections due to finite energy are relevant for strong field, it would be interesting, and perhaps more appropriate, to study their effects in the derivation of general relativity also given in \cite{Verlinde:2010hp}.  

In the following section, we discuss the physical implications of atoms with maximal energy, without referring to the details of the microscopic physics. 
In Sec.\ \ref{se_mod}, we discuss a simple explicit model of the atoms. A discussion of the results can be found in Sec.\ \ref{se_conc}. 

\section{Minimal radius from maximum energy}
\label{se_rmin}
Let us first recapitulate Verlinde's derivation of Newton's law, in a suitably generalized form. For this, we assume an unspecified relation $T=T(E_\at)$ between the temperature and the average energy of an atom of area. 

We consider a spherical equipotential surface with area $A=4\pi R^2$ around a point mass $M$. Then the gravitational acceleration of a test mass at distance $R$ is calculated as 
\begin{align}
a&=2\pi\frac{ck_B}{\hbar}T(E_\at)\\
&=2\pi\frac{ck_B}{\hbar}T(Mc^2/N_\at)\\
&=2\pi\frac{ck_B}{\hbar}T\left(\frac{1}{4\pi}c^2l^2_\at \frac{M}{R^2}\right).
\label{eq_newt}
\end{align}
For the usual energy equipartition $E=\alpha k_\text{B} T$ with $\alpha$ a numerical constant, we get 
\begin{equation}
a= \frac{1}{2\alpha}\frac{c^3l^2_\at}{\hbar}\frac{M}{R^2},
\end{equation}
and are led to identify
\begin{equation}
l_\at^2=2\alpha \lp^2. 
\end{equation}
This is essentially the argument by Verlinde, with $\alpha$ taking into account different prefactors in the equipartition law (see \cite{Klinkhamer:2011gu,Neto:2011wg} for a discussion and possible consequences of such a prefactor).   

When considering systems with maximum energy, the relation between average energy $E_\at$ and temperature is very different. There is a highest energy that can be attained, $E_\at\leq E_\ma$, and so, after some approximately linear growth, the average energy per atom saturates (see Fig.\ \ref{fi_generic}). 
This has important consequences: For a derivation of Newton's law as above, 
$M$ and $R$ must be independent, but this is no longer true in general. Since 
\begin{equation}
\frac{E}{N_\at}=E_\at \leq E_\ma
\end{equation}
and $N_\at$ is related to $A$ via \eqref{eq_nat}, and $A=4\pi R^2$, we find a minimal value  
\begin{equation}
R_\mi
=\frac{1}{\sqrt{4\pi}} \sqrt{\frac{E}{E_\ma}}l_\at 
=\sqrt{\frac{\alpha}{2\pi}} \sqrt{R_\text{S}R_\at}
\label{eq_rmin}
\end{equation}
for $R$. Here $R_\at= c \hbar/E_\ma$ is the (reduced) Compton wavelength of an atom at maximum energy. 

\section{A simple model}
\label{se_mod}
To illustrate the statements of the preceding section, let us consider a very simple model for the atoms of area: We assume them to have $N$ nondegenerate equidistant energy levels with  energy spacing $E_0$. The Hamiltonian is thus given by
\begin{equation}
H=E_0\sum_{n=1}^{N} n \ket{n}\bra{n} 
\end{equation}
with the energy eigenstates $\ket{n}$. We consider the system immersed in a heat bath of inverse temperature $\beta=1/(k_BT)$. With $\rho_\beta=\exp(-\beta H)$ and $Z(\beta)=\tr\rho_\beta$, one finds   
\begin{align}
E_\at:=&\expec{H}_\beta\\
&=-\frac{\text{d}}{\text{d}\beta}\ln Z(\beta)\\
&=\frac{E_0 \left(e^{\beta  E_0 (N+1)}-(N+1) e^{\beta  E_0}+N\right)}{\left(e^{\beta 
   E_0}-1\right) \left(e^{\beta  E_0 N}-1\right)}.\label{eq_exact}
\end{align}
For large temperatures, this has a well defined limit 
\begin{equation}
E_\ma=\lim_{T\rightarrow\infty}E_\at(T)=\frac{1}{2}(N+1)E_0.
\end{equation}
In Fig.\ \ref{fi_generic} we have plotted $E_\at(T)$ for the case $N=10$. A linear regime is visible, with corrections for low $T$ due to nonvanishing 
$E_0$, and saturation for large $T$.  

\begin{figure}
\centerline{\epsfig{file=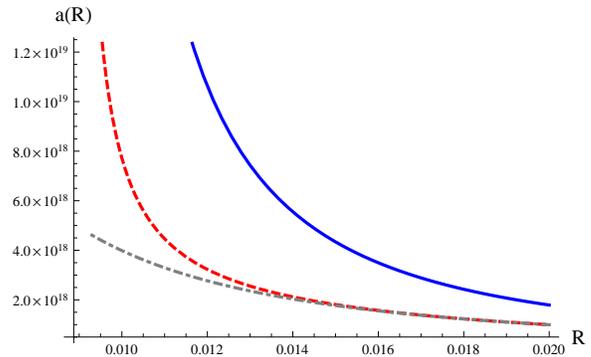,scale=0.6}}
\caption{\label{fi_div} Acceleration versus radius in SI units, with $M=M_\text{Earth}$ and $E_\ma$ such that $R_\at= R_\sch$: General relativity (solid curve), nonrelativistic entropic gravity (dashed curve), Newton's law (dot-dashed curve)} 
\end{figure}

To separate off the corrections at low temperature, which we will not further discuss here, and to bring out the linear regime, it is useful to take the limit $N\rightarrow \infty, E_0\rightarrow 0$ with $E'_\ma:=NE_0$ \emph{fixed}. 
In this limit
\begin{equation}
E_\at(T)=k_\text{B}T-\frac{E'_\ma}{\exp(E'_\ma/k_\text{B}T)-1}.
\label{eq_lim}
\end{equation}
\begin{figure}
\centerline{\epsfig{file=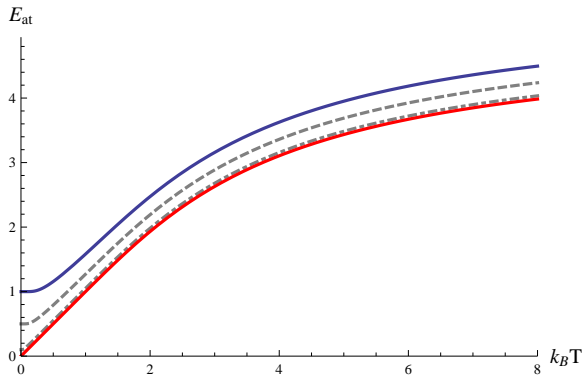,scale=0.6}}
\caption{\label{fi_lim} Energy versus temperature in units of some fixed energy $E$, for $E_0=E, N=10$ (upper solid curve) $E_0=E/5, N=50$ (dashed curve) $E_0=E/10, N=100$ (dot-dashed curve) and for \eqref{eq_lim}, with 
$E'_\ma=10E_0$.} 
\end{figure}
This is a very good approximation to the actual high energy behavior (up to a constant shift in energy), as can, for example, be seen from Fig.\ \ref{fi_lim}. 
The first term is the linear part, the second term a correction that becomes dominant at large $T$ and leads to a maximum average energy per atom of 
$E_\ma=E'_\ma/2$. Thus there is indeed a linear regime as needed for the derivation of Newton's law, provided $E_\ma$ is large enough. There will be corrections for high $T$, i.e., for small radius. Let us also note the scaling behavior of the energy-temperature relation: $E_\at/E_\ma$ is solely a function of the ratio $T/E_\ma$. Thus we also have 
\begin{equation}
T\equiv T(E_\at,E_\ma)=E_\ma f(E_\at/E_\ma). 
\end{equation}
Neither \eqref{eq_exact}, nor \eqref{eq_lim}, can be inverted (below $E_\ma$) in terms of elementary functions, so we can not display these corrections explicitly, but we will 
give two illustrative examples. First we consider gravitational acceleration at the surface of the Earth. The associated Unruh temperature is  
\begin{equation}
T=\frac{1}{2\pi}\frac{\hbar}{ck_B}g\approx 4.0 \times 10^{-20} \text{ K}, 
\end{equation} 
corresponding to an energy
\begin{equation}
k_BT\approx 5.5\times 10^{-43}\text{ J}\approx 3.4 \times 10^{-24} \text{ eV}. 
\end{equation}
Now we assume $E_\ma\simeq 1$eV. Then relative corrections to Newton's law at the surface of the Earth would be extremely tiny, of the order of $\exp(-10^{24})$, and thus completely unmeasurable. 
Also, $R_\at\simeq 2\times 10^{-7}\text{m}$ is much smaller than $R_\sch\simeq 9\times 10^{-3} \text{m}$, so corrections to Newton's law would still be imperceptible near a black hole of earth mass. 
If on the other hand $E_\ma\simeq 10^{-5}$eV, the relative corrections to Newton's law at the surface of the Earth would still be extremely tiny, of the order of $\exp(-10^{19})$, but then 
$R_\at\simeq R_\sch$, so the gravitational acceleration would diverge at the Schwarzschild radius. Note however, that there are reasons to believe that $E_\ma$ has to be very high, see the note added at the end of Sec.\ \ref{se_conc}.  We have plotted the gravitational acceleration obtained in this way, together with the prediction from general relativity for a static observer,   
\begin{equation}
a=\frac{1}{1-\frac{2MG}{c^2 R}}\frac{MG}{R^2},
\label{eq:}
\end{equation}
and from Newton's law in Fig.\ \ref{fi_div}. 
One sees how the entropic gravity curve first tracks Newton's law, then bends to join general relativity on the horizon. We remind the reader that $R_\at\simeq R_\sch$ in this example is due to the fine-tuning of the parameters. With $E_\ma$ remaining fixed, 
the gravitational acceleration would diverge -- for different masses --  at radii larger or smaller than the Schwarzschild radius. 
\section{Conclusions}
\label{se_conc}
In the present work we have studied the effects that the assumption of atoms of area of bounded energy has for the derivation of Newton's law in Verlinde's entropic gravity.
Motivating this assumption is the finiteness of black hole entropy: It is natural to identify this entropy with the one from the entropic gravity scenario for the special case that the equipotential surface is a black hole horizon. This suggests that the atoms of area have a finite dimensional state space in general, and hence bounded energy. 

We saw that the assumption of  bounded energy led, intriguingly, to the divergence of the acceleration at some nonzero radius $R_\mi$. We could, however, not identify this phenomenon straightforwardly with the divergence of the acceleration of a static observer at the Schwarzschild radius, since we found that $R_\mi$ scales in a way that involves both, the mass $M$ of the body, and the energy scale $E_\ma$ of the microscopic physics. Identification of $R_\sch$ with $R_\mi$ could only be accomplished if the maximum energy $E_\ma$ of the atoms that make up the holographic screen would depend on the mass of the body, in a specific way. This however, is not very plausible given the physical picture of entropic gravity. 

Implicit in the use of the equipartition law, and thus also in our present considerations, is the assumption that the degrees of freedom of the screen are immersed in a heat bath of a certain temperature $T$, given by \eqref{eq_unruh}. It is not clear which physical system constitutes this heat bath. It stands to reason that it can not just be the Unruh radiation that an accelerated particle is immersed in, since space itself is supposed to emerge from a course-graining of the degrees of freedom \cite{Verlinde:2010hp}. Thus this point merits further investigation. 

What are the consequences of our results for entropic gravity? First of all, it can be the case that the energy of the atoms of the holographic screen is not finite. After all, only entropy changes enter the derivation of Newton's law; the entropy itself may be infinite. In fact, entropy can still be finite for an infinite-dimensional system, if taken in a suitable state. \emph{If} the atoms of area do have bounded energy, our arguments show that Verlinde's reasoning is certainly viable, as long as the maximum energy is not too low. Furthermore we see no physical reason to suggest that this maximum energy has to be low. In fact, there are reasons to believe that it has to be very high; see the note added at the end of this section. If the maximum  energy is not too high, there could be interesting phenomenological consequences for strong gravitational fields. It should however be said that the entire picture of entropic gravity should be regarded with caution in the strong field regime, as 
new physical effects may enter the stage.

\paragraph*{Note added.} We would like to add the following very useful observation by one of the referees regarding the energy scale $E_\ma$. This scale must certainly be high enough for the screen to be able to hold the mass-energy that appears to be behind the screen. The ratio of mass-energy per screen area is highest for the horizon of a black hole, and turns out to be inversely proportional to the Schwarzschild radius in this case. A lower bound 
\begin{equation}
E_\ma\geq \frac{c\hbar}{8\pi R_{\text{S,min}}} 
\end{equation}
can be given in terms of the Schwarzschild radius $R_{\text{S,min}}$ of the smallest physically possible black hole. Even for conservative assumptions about $R_{\text{S,min}}$, the resulting bound is very high. Thus corrections for astrophysical black holes seem to be completely negligible. Moreover, an upper bound on the minimal radius from entropic gravity is then given by 
\begin{equation}
R_\mi\leq \sqrt{4\alpha}\sqrt{R_\text{S} R_{\text{S,min}}}.
\end{equation}
$R_\mi$ is thus always smaller than (or at least of the same order of magnitude as) the Schwarzschild radius. Corrections are only appreciable very close to extremely microscopic black holes. 
This argument assumes that the area atoms have universal properties and there is no binding energy between them. The former is a reasonable assumption, and the latter was assumed already in Verlinde's derivation. The apparently very high $E_\ma$, together with the observation that the dimensionality of the internal state space of the atoms seems to have to be quite small \cite{Klinkhamer:2010qa}, seems to lead to a tension with Newton's law for weak fields, as a coarsely spaced energy spectrum would lead to corrections in this regime. It may be interesting to further study the implications.  

\begin{acknowledgments}
The author thanks F.R.\ Klinkhamer for helpful discussions on various aspects of entropic gravity, and for useful comments on a draft of the present work. Comments by one of the referees are greatly appreciated. They led to various improvements of the manuscript, in particular, to the note added at the end of Sec.\ \ref{se_conc}.
The research was partially supported by the Spanish MICINN Project No.\ FIS2008-06078-C03-03.
\end{acknowledgments}

\end{document}